\documentclass[10pt,twocolumn,a4paper]{article}

\usepackage[margin=1.8cm,columnsep=0.6cm]{geometry}
\usepackage[utf8]{inputenc}
\usepackage[T1]{fontenc}
\usepackage{amsmath,amssymb}
\usepackage{graphicx}
\usepackage{booktabs}
\usepackage{array}
\usepackage[font=small,labelfont=bf]{caption}
\usepackage{xcolor}
\usepackage{cite}
\usepackage{authblk}

\usepackage[colorlinks=true,linkcolor=blue!60!black,citecolor=blue!60!black,urlcolor=blue!60!black]{hyperref}

\newcommand{\LZ}{Ref.~\cite{LZ:2026ext}}

\title{The LZ event with a two-state dark matter halo}

\author[1,2,3]{Vin\'icius Oliveira}
\affil[1]{Departamento de F\'isica da Universidade de Aveiro, Campus de Santiago, 3810-183 Aveiro, Portugal}
\affil[2]{Laborat\'orio de Instrumenta\c{c}\~ao e F\'isica Experimental de Part\'iculas (LIP), Universidade do Minho, 4710-057 Braga, Portugal}
\affil[3]{Lund University, SE-223 62 Lund, Sweden}

\date{}

\begin{document}

\twocolumn[
  \begin{@twocolumnfalse}
    \maketitle
    \begin{abstract}
    The LUX-ZEPLIN (LZ) experiment has reported a single nuclear-recoil candidate at $E_R \simeq 248$~keV. LZ interprets it in terms of inelastic dark matter that scatters endothermically from its ground state, $\chi_1 N \to \chi_2 N$. In pseudo-Dirac models the excited state $\chi_2$ can survive to the present day, and exothermic scattering, $\chi_2 N \to \chi_1 N$, is then also open. The published LZ intervals do not describe this case. We rebuild the LZ statistical inference  and validate it against the published intervals. We then apply it to a halo that contains a fraction $f$ of $\chi_2$. For $f \gtrsim 10^{-5}$--$10^{-2}$, depending on the mass splitting, the exothermic contribution dominates. For $f=1/2$  the cross section required by the candidate drops by factors of $80$ to $1.6\times10^4$ for $\delta = 250$--$350$~keV, and the kinematic cutoff of the endothermic signal disappears. A benchmark that explains the event with $\chi_1$ alone lies a factor of $30$ above the upper limit when half of the halo is $\chi_2$.
    \end{abstract}
    \vspace{1em}
  \end{@twocolumnfalse}
]

\section{Introduction}
\label{sec:intro}

The nature of dark matter (DM) remains unknown~\cite{Bertone:2004pz}. Its gravitational effects are established on many scales, from galactic rotation curves to the cosmic microwave background, where it accounts for about $26\%$ of the energy density of the Universe~\cite{Planck:2018vyg}. Whether DM also interacts with ordinary matter is still open. Direct-detection experiments address this question by searching for nuclear recoils induced by DM scattering. Liquid-xenon detectors set the strongest constraints on elastic, spin-independent scattering above a few GeV~\cite{LZ:2024ws}.

LUX-ZEPLIN (LZ) has recently extended its nuclear-recoil window to $270$~keV~\cite{LZ:2026ext}. In $2.84$~t\,yr of exposure it observed one event consistent with a nuclear recoil (NR) of $E_R = 248 \pm 23\,({\rm stat}) \pm 23\,({\rm sys})$~keV, in a region where the expected background is about $0.01$ events. The collaboration tested $616$ interaction models and found a maximum local significance of $3.4\sigma$, which becomes $2.6\sigma$ after the look-elsewhere correction. The event is not a discovery. It is, however, unusual. Elastic spin-independent scattering explains it poorly, since a spectrum normalized to one event near $250$~keV predicts many more events at low energy, where none are seen.

The event has prompted several papers. The largest group invokes inelastic DM~\cite{TuckerSmith:2001hy}. If the DM particle $\chi_1$ must be excited into a heavier state $\chi_2$ to scatter, the mass splitting $\delta$ sets a velocity threshold, suppresses low-energy recoils and moves the spectrum to high energy. LZ itself tested this endothermic case, $\chi_1 N \to \chi_2 N$, through the operator $\mathcal{O}_1$ of the non-relativistic effective theory~\cite{Fitzpatrick:2012ix,Anand:2013yka}, with $\delta \leq 350$~keV. Several works identify the relevant region of $(m_\chi, \delta)$ at the kinematic edge of the endothermic signal~\cite{Su:2026ine,DiMauro:2026lz,Visinelli:2026pq}. Others embed the splitting in concrete models: singlet-doublet fermions~\cite{Borah:2026zwf}, dark photons~\cite{Zhu:2026dag}, mixing-suppressed and asymmetric inelastic DM~\cite{Lee:2026jxl,Nagata:2026pbj}, transition magnetic dipoles~\cite{He:2026hqz}, inelastic scalars in extended gauge sectors~\cite{Kumar:2026lgi,Das:2026buc}, radiative neutrino-mass models~\cite{Borah:2026ris}, sneutrinos~\cite{Lian:2026hpm}, axion portals~\cite{An:2026pkc}, hadrophilic freeze-in~\cite{He:2026idw}, and pseudo-Dirac fermions produced by freeze-in at low reheating temperature~\cite{CaboAlmeida:2026fi}. Higgsino DM belongs to the same class, since its two neutral states are split by mixing with the gauginos~\cite{Freese:2026hig,Fan:2026hig,Wu:2026hig,Cheung:2026byg,Langhoff:2026ujr}, but it is now strongly constrained by the high-energy sideband~\cite{Safdi:2026sb}, by solar capture~\cite{Pospelov:2026sun,Ghosh:2026txe} and by neutrino telescopes~\cite{Bose:2026ndd}. Related studies compare nuclear and dark-sector excitations~\cite{Khan:2026nwp} and discuss xenon excitation signals~\cite{Gu:2026vto}.

A second group reverses the transition. If the relic is the heavier state, the exothermic scattering $\chi_2 N \to \chi_1 N$~\cite{Graham:2010ca} releases the splitting to the nucleus. The spectrum then peaks at a recoil energy fixed by $\delta$, and the whole halo contributes. This has been proposed with $\chi_2$ as the relic~\cite{Baer:2026exo,deLima:2026exo}, together with a Galactic Center excess~\cite{Gemmell:2026lz}, and as a benchmark against argon~\cite{Baer:2026ar}. Ref.~\cite{Fan:2026ine} studies both transitions, but separately. A third group avoids the standard halo altogether. It includes DM boosted by cosmic rays or other mechanisms~\cite{Liang:2026coz,Alhazmi:2026efz,Heikinheimo:2026kwp,Mahapatra:2026glu,Chauhan:2026udz}, a high-velocity halo component~\cite{OHare:2026nqi}, absorption~\cite{Lou:2026abs}, neutron disappearance~\cite{Aghaie:2026vsu,Lee:2026zbr,Uttayarat:2026isp}, up-scattered atmospheric neutrinos~\cite{Jeesun:2026vzo}, and models with vector-like leptons, warped extra dimensions, non-standard Yukawa couplings or dark QCD~\cite{Elahi:2026vlm,Lee:2026xxh,De:2026win,Sannino:2026hkc,Arcadi:2026kev}.

These works confront the data in simplified ways. Most normalize the predicted spectrum to one event~\cite{Visinelli:2026pq,Baer:2026exo,CaboAlmeida:2026fi}. Others use an energy-only likelihood with a sideband~\cite{deLima:2026exo,Gemmell:2026lz,Mahapatra:2026glu}, or two or three Poisson bins in energy~\cite{Su:2026ine,Fan:2026ine}. To our knowledge, none of them uses the LZ background model or the discrimination between electron and nuclear recoils, and where intervals are given they follow from asymptotic formulas.

The inelastic interpretations share a feature that the analyses above do not capture. The splitting requires two nearly degenerate states, typically a pseudo-Dirac pair, and both are produced in the early Universe. If $\chi_2$ is long-lived, as in some realizations cited above, the present-day halo contains both states, and the signal is the sum of endothermic and exothermic scattering. The intervals published by LZ assume a halo of $\chi_1$ alone, so they do not apply to this case. Neither do the simplified confrontations above, which treat the two transitions separately or not at all.

In this Letter we address this gap by rebuilding the LZ statistical inference from public information, validate it against the published intervals, and applying it to a halo that contains a fraction $f$ of $\chi_2$\footnote{Throughout this Letter we use the Standard Halo Model with the parameters of Ref.~\cite{Baxter:2021pqo}, which LZ follows: $\rho = 0.3$~GeV\,cm$^{-3}$, $v_0 = 238$~km\,s$^{-1}$, $v_{\rm esc} = 544$~km\,s$^{-1}$, and the year-averaged Earth velocity, $|v_E| = 253.7$~km\,s$^{-1}$.}.

\section{Model and kinematics}
\label{sec:model}

We consider a Dirac fermion charged under a dark U$(1)$ with a massive gauge boson $Z'$~\cite{DiMauro:2026lz}. Small Majorana masses split it into two Majorana states, $\chi_1$ and $\chi_2$, with $\delta \equiv m_{\chi_2} - m_{\chi_1}$. The vector current of a Majorana fermion vanishes, so the $Z'$ coupling is off-diagonal, $i g_\chi Z'_\mu \bar\chi_2\gamma^\mu\chi_1$. Every scattering off a nucleus changes the state. Once the $Z'$ is integrated out, the quark vector current gives the isoscalar coupling $c_1^s = 3 g_\chi g_q/m_{Z'}^2$ to $\mathcal{O}_1 = 1_\chi 1_N$, the operator $\mathcal{O}_1^s$ of LZ, with DM--nucleon cross section
\begin{equation}
\sigma_N = \mu_N^2 (c_1^s)^2/\pi .
\label{eq:sigmaN}
\end{equation}

Both thermal and non-thermal production leave comparable abundances of the two states. In freeze-in, the off-diagonal coupling produces them in pairs, ${\rm SM} \;  {\rm SM} \to \chi_1\chi_2$, one of each per event. In thermal freeze-out, the two states decouple at $T \sim m_\chi/20 \gg \delta$, when their ratio is $e^{-\delta/T} \simeq 1$. Unless $\chi_2$ later decays or converts into $\chi_1$, it therefore makes up about half of the DM today. We define the fraction of the local density in the excited state, $f = \rho_{\chi_2}/\rho$.

The differential scattering rate per unit detector mass is \cite{Lewin:1995rx}
\begin{multline}
\frac{dR}{dE_R} = \frac{\rho\,\sigma_N}{2 m_\chi \mu_N^2}\sum_i \xi_i A_i^2 F_i^2(E_R)\\
\times\Big[(1-f)\,\eta\big(v_{\min,i}^{+}\big) + f\,\eta\big(v_{\min,i}^{-}\big)\Big],
\label{eq:rate}
\end{multline}
with $\xi_i$ the mass fractions of the xenon isotopes, $\eta$ the mean inverse speed, and
\begin{equation}
v_{\min}^{\pm} = \frac{|\,m_A E_R/\mu_{\chi A} \pm \delta\,|}{\sqrt{2 m_A E_R}} .
\label{eq:vmin}
\end{equation}
The $+$ sign represents an endothermic process, and the $-$ one exothermic. The cross section is common to both, since the reduced masses differ by a relative amount of order $10^{-7}$.

The two processes (endothermic and exothermic) behave in opposite ways. Both cases have a special recoil energy, $E_R^\star = \mu_{\chi A}\delta/m_A$. In the endothermic case it is where $v_{\min}^+$ is smallest, $v_{\min}^{+\star} = \sqrt{2\delta/\mu_{\chi A}}$. Only the fastest particles scatter, and the signal vanishes above
\begin{equation}
\delta_{\max} = \tfrac12\,\mu_{\chi A}\big(v_{\rm esc} + |v_E|\big)^2 \,.
\label{eq:deltamax}
\end{equation}
In the exothermic case $v_{\min}^- = 0$ at $E_R^\star$. Even a particle at rest produces a recoil of energy $E_R^\star$. The whole halo contributes, and there is no kinematic cutoff.

\begin{figure*}[t]
\centering
\includegraphics[width=0.95\textwidth]{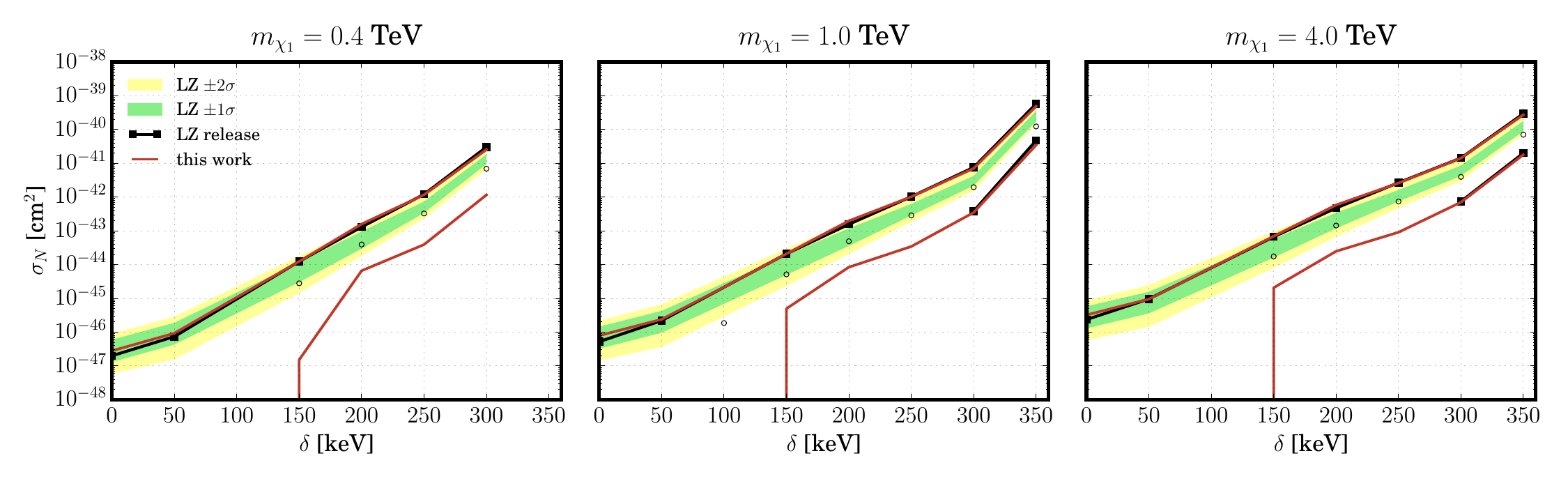}
\caption{90\% C.L. intervals for endothermic $\mathcal{O}_1^s$. Black: LZ data release~\cite{LZ:2026hepdata}, with the $\pm1\sigma$ and $\pm2\sigma$ sensitivity bands. Red: this work, circles mark the best fit.}
\label{fig:release}
\end{figure*}

\section{The recast}
\label{sec:recast}

The rate in Eq.~\eqref{eq:rate} combines three inputs: the particle physics of the model ($m_\chi$, $\delta$, $f$ and $\sigma_N$), the halo through $\eta(v_{\min})$, and the nuclear form factor $F^2(E_R)$. Near the candidate, the form factor is the most important of them. It has diffraction minima near $E_R = 100$ and $265$~keV, and the candidate lies just below the second one. Most phenomenological studies use the Helm form factor~\cite{Helm:1956zz,Lewin:1995rx}. At $248$~keV, it overestimates the shell-model response used by LZ by a factor of $5$. We therefore use the shell-model response of Refs.~\cite{Fitzpatrick:2012ix,Anand:2013yka}, as described in the Supplemental Material.

The LZ Collaboration computes the nuclear response with WimPyDD~\cite{Jeong:2021bpl} and DMFormfactor-v6~\cite{Anand:2013yka}. We take the shell-model response of xenon from \texttt{dmdd}~\cite{Gluscevic:2015sqa,dmdd}, split it among isotopes with the Helm radii, and multiply it by the ratio $C(E_R)$ between the LZ spectrum at $\delta = 0$ (Fig.~1 of \LZ) and ours. For $\mathcal{O}_1^s$ the rate factorizes into a nuclear part, which depends on $E_R$ only, and a halo part. $C(E_R)$ at $\delta =0$  fixes the nuclear response to that of LZ for any $\delta$ and $m_\chi$. It is an input to the signal model, not a fit to the limits. With it, our spectra at $\delta = 200$ and $300$~keV agree with those of LZ within 9\% at $248$~keV (Supplemental Material).

The LZ Collaboration evaluates each event in $(S1c,$ $ \log_{10}S2c)$ with an unbinned likelihood. The events are public~\cite{LZ:2026hepdata}, but at the time of writing, the signal and background densities are not. The unbinned likelihood therefore cannot be rebuilt from the outside. Instead, we build a binned likelihood from the public projection of Fig.~5 of \LZ, which gives the same densities integrated over bins: three slices $k$ in $S1c$ ($<250$, $250$--$500$ and $>500$~phd), each divided into $32$ bins $j$ of width $0.5$ in the distance to the NR band\footnote{Since data, background and signal share the same binning in $(\log_{10}S2c - \mu_{\rm NR})/\sigma_{\rm NR}$, we never need $\mu_{\rm NR}$ or $\sigma_{\rm NR}$.}, $(\log_{10}S2c - \mu_{\rm NR})/\sigma_{\rm NR}$. In each bin it gives the observed counts $n_{kj}$, the post-fit background $b_{kj}$, and the shape of an NR signal. 

Our binned likelihood is defined as
\begin{equation}
\mathcal{L}(s) = \prod_{k=1}^{3}\prod_{j=1}^{32} \frac{\mu_{kj}^{\,n_{kj}}}{n_{kj}!}\,e^{-\mu_{kj}},
\quad
\mu_{kj} = s\,F_k P_{kj} + b_{kj},
\label{eq:likelihood}
\end{equation}
where $s = \mathcal{E}\int dE_R\,\varepsilon(E_R)\,dR/dE_R$ is the number of signal events, with $\mathcal{E} = 2.84$~t\,yr and $\varepsilon$ the efficiency of Fig.~S2 of \LZ. 

$F_k$ is the fraction of the signal in slice $k$. We compute it from the efficiency-weighted differential rate, $\varepsilon(E_R)\,dR/dE_R$, of Eq.~\eqref{eq:rate}. The slices are defined in $S1c$, so we convert their edges to recoil energy using the constant-energy contours of Fig.~4 of \LZ. The NR median crosses the two edges at $E_R = 128.7$ and $232.3$~keV, which we take as the energy boundaries of the slices. Applied to the $\mathcal{L}_{10}^s$ spectrum of LZ, this cut reproduces the slice fractions of Fig.~5 within 1\%. $P_{kj}$ is the distribution of the signal in $(\log_{10}S2c - \mu_{\rm NR})/\sigma_{\rm NR}$ within slice $k$. It is set by the detector response to a nuclear recoil, which depends on the recoil energy and not on the interaction that produced it. LZ passes all signals through the NEST model. We therefore take $P_{kj}$ from the $\mathcal{L}_{10}^s$ signal of Fig.~5. The background is fixed at its post-fit value. LZ instead profiles the normalization of each component, but Table~I of \LZ\ shows that the fit moves every component by at most $1\sigma$ from its expected value. In the highest slice, the background totals $0.01$ events, so its variations have a negligible effect on the result.

The test statistic is $\tilde t_s = -2\ln[\mathcal{L}(s)/\mathcal{L}(\hat s)]$, with $\hat s \geq 0$. Following LZ, we build the distribution of $\tilde t_s$ from $3000$ toy Monte Carlo datasets per value of $s$~\cite{Baxter:2021pqo}. Each toy draws Poisson counts in the $96$ bins, $n_{kj}\sim{\rm Pois}(sF_kP_{kj}+b_{kj})$, with the background fixed at its post-fit value. The $p$-value of $s$, $p(s)$, is the fraction of these toys with $\tilde t_s$ at least as large as observed in the data. The 90\% C.L. interval is a Neyman construction in a raster scan over $\delta$: it contains all values of $s$ with $p(s) \geq \alpha$, where $\alpha = 1 - {\rm C.L.} = 0.10$.

The local significance, $Z$, quantifies how incompatible the data are with the background-only hypothesis, $s = 0$. We compute $\tilde t_0$ on the data and on $10^5$ toys generated with $s = 0$, i.e.\ $n_{kj}\sim{\rm Pois}(b_{kj})$. The $p$-value, $p_0$, is the fraction of toys with $\tilde t_0$ above the observed value, and $Z = \Phi^{-1}(1-p_0)$, with $\Phi$ the standard normal cumulative distribution. This toy-based $p_0$ replaces the asymptotic formula, $Z = \sqrt{\tilde t_0}$~\cite{Cowan:2010js}, which assumes that $\tilde t_0$ follows a $\chi^2$ distribution. That assumption holds only when the relevant bins contain many events. Here the bins where the signal concentrates are nearly empty: the candidate bin expects $3.7\times10^{-5}$ background events. A background-only toy rarely places an event there, so large values of $\tilde t_0$ are much rarer than the $\chi^2$ predicts. The asymptotic formula therefore overestimates $p_0$. At $\delta = 150$~keV, it gives $p_0 = 0.11$, while the toys give $0.07$.

\begin{figure*}[!ht]
\centering
\includegraphics[width=0.95\textwidth]{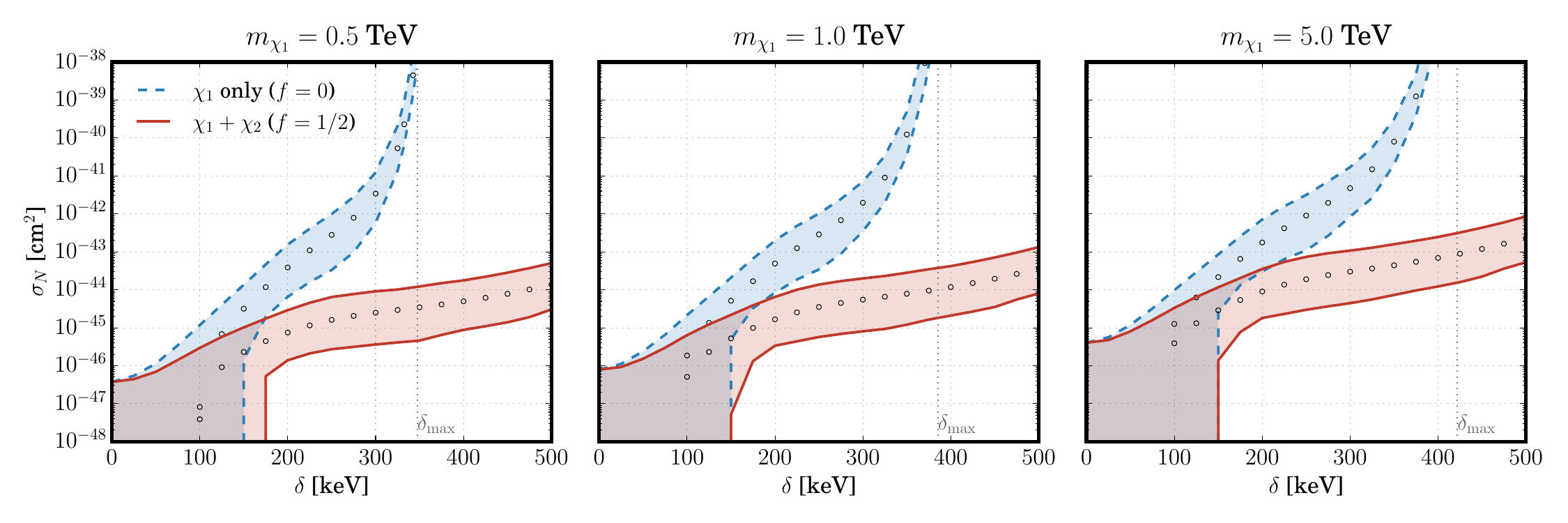}
\caption{90\% C.L. intervals for $\chi_1$ only ($f = 0$, blue, dashed) and $\chi_1 + \chi_2$ with equal abundances ($f = 1/2$, red, solid). Circles mark the best fit. The dotted line is the endothermic cutoff, Eq.~\eqref{eq:deltamax}. The dotted line is the kinematic endpoint of the endothermic signal, Eq.~\eqref{eq:deltamax}. The blue region diverges earlier, because near the endpoint the recoils lie at $E_R^\star = \mu_{\chi A}\delta/m_A$, above the LZ window for $m_{\chi_1} \gtrsim 1$~TeV.}
\label{fig:masses}
\end{figure*}

Figure~\ref{fig:release} compares our intervals with the LZ data release~\cite{LZ:2026hepdata} at the three published masses. For $\delta \geq 50$~keV the upper limits agree within 26\% at all $17$ points, and within 5\% at about half of them (Table~\ref{tab:upper}). Where the release gives a lower limit, ours lies within 27\% of it.  Two deviations remain. At $\delta = 0$ our upper limits are 40--50\% weaker, because $P_{kj}$ in the lowest slice is not representative of a signal below $50$~keV. Additionally, our local significances are $0.2$--$0.7\sigma$ below those of LZ. At $1$~TeV we find $Z = 1.5$, $2.3$ and $2.5$ at $\delta = 150$, $250$ and $300$~keV, against $2.2$, $2.9$ and $3.0$. The most likely cause is the binning: our likelihood uses only three $S1c$ slices and loses the energy information inside each of them, which the unbinned likelihood of LZ retains.

\begin{table}[t]
\centering\footnotesize\setlength{\tabcolsep}{3.2pt}
\begin{tabular}{cccccccc}
\toprule
$m_\chi$ [GeV] & $\delta = 0$ & $50$ & $150$ & $200$ & $250$ & $300$ & $350$ \\
\midrule
400  & 1.40 & 1.25 & 1.00 & 1.22 & 0.97 & 0.85 & -- \\
1000 & 1.51 & 1.09 & 1.00 & 1.26 & 0.99 & 0.94 & 0.84 \\
4000 & 1.37 & 0.99 & 1.04 & 1.23 & 0.97 & 0.99 & 0.92 \\
\bottomrule
\end{tabular}
\caption{Ratio of our 90\% C.L. upper limit on $\sigma_N$ to that of the LZ release; $\delta$ in keV.}
\label{tab:upper}
\end{table}

The lower limit follows directly from the local significance. At $s = 0$, the test statistic $\tilde t_0$ is the discovery statistic, and $p_0$ is the background-only $p$-value. The Neyman interval therefore excludes $s = 0$, i.e.\ has a non-zero lower limit, if and only if
\begin{equation}
p_0 < \alpha
\quad\Longleftrightarrow\quad
Z > \Phi^{-1}(1-\alpha) = 1.28 \,.
\label{eq:onset}
\end{equation}
Our recast follows this relation. At $1$~TeV, our $Z$ crosses $1.28$ at $\delta = 140$~keV, and our lower limit appears at the same splitting (Fig.~\ref{fig:onset} in the Supplemental Material).

Applied to the LZ significances, Eq.~\eqref{eq:onset} places the onset near $\delta = 117$~keV. This value is a consistency check of the LZ results, not an input to our recast. Our intervals and significances both come from our binned likelihood, so our onset follows our own $Z$. Our onset lies later because our significances are lower, a consequence of the binned likelihood. The published LZ results depart from Eq.~\eqref{eq:onset} in both directions. Figure~S7 of \LZ\ shows a lower limit already at $\delta = 100$~keV, where $Z = 0.8$. The data release, instead, gives a non-zero lower limit only at the four points with $Z \geq 3.0$, which at $1$~TeV means $\delta \geq 300$~keV. We have not identified the origin of these differences. They may reflect details of the LZ implementation not described in \LZ, such as the treatment of nuisance parameters, the grid in the coupling, or a reporting convention of the release. The release also has no entry at $\delta = 100$~keV, and its upper limits differ from those of Fig.~S7 by up to $13\%$.

\begin{figure}[t]
\centering
\includegraphics[width=\columnwidth]{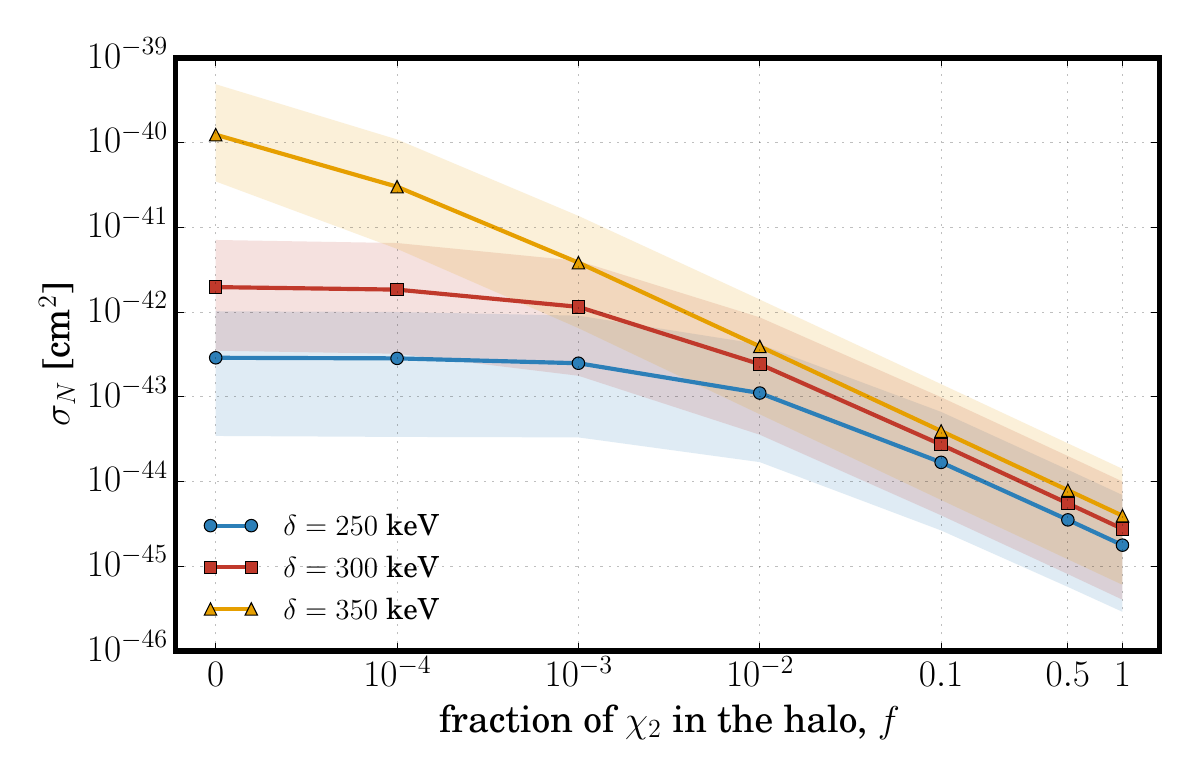}
\caption{Best fit (lines) and 90\% C.L. interval (bands) on $\sigma_N$ versus the fraction $f$ of $\chi_2$ in the halo, at $m_{\chi_1} = 1$~TeV. The leftmost point is $f = 0$.}
\label{fig:fraction}
\end{figure}
\section{Results}
\label{sec:results}

The two-component signal enters the same likelihood. Only $F_k$ changes, since $P_{kj}$, $b_{kj}$ and $n_{kj}$ describe the detector and the data. Figure~\ref{fig:masses} shows the intervals for $f = 0$ and $f = 1/2$ at three masses. LZ publishes intervals only at $0.4$, $1$ and $4$~TeV.

At $1$~TeV the required cross section drops by a factor of $80$ at $\delta = 250$~keV, $360$ at $300$~keV and $1.6\times10^4$ at $350$~keV. It becomes nearly independent of $\delta$, from $3.5\times10^{-45}$ to $3.6\times10^{-44}$~cm$^2$ between $250$ and $500$~keV, and the endothermic cutoff disappears. The reason is kinematic. The exothermic case reaches $v_{\min} = 0$ near the candidate, and the whole halo contributes, not only its tail. The same pattern holds at all masses (Table~\ref{tab:masses}). For $f = 1/2$ the region extends to $500$~keV\footnote{For $f = 1/2$, we impose no upper bound on $\delta$. A population of $\chi_2$ in the halo today requires, however, that its lifetime exceeds the age of the Universe. Larger splittings may open faster decay channels, $\chi_2 \to \chi_1 \gamma$ and $\chi_2 \to \chi_1 \nu\bar\nu$ with rates that grow as powers of $\delta$, and $\chi_2 \to \chi_1 e^+e^-$ above $\delta = 2m_e \simeq 1$~MeV. A sufficiently large $\delta$ would therefore deplete $\chi_2$ and set $f \to 0$. This bound is model dependent, and we do not include it here.}, and the required cross section grows approximately linearly with $m_{\chi_1}$, as expected from the number density once the kinematics no longer restricts the signal.

\begin{table}[t]
\centering\footnotesize\setlength{\tabcolsep}{2.4pt}
\begin{tabular}{lccccc}
\toprule
$m_{\chi_1}$ [TeV] & 0.5 & 1 & 2.5 & 5\\
\midrule
$f = 0$ & $3.4\times 10^{-42}$ & $2.0 \times 10^{-42}$ & $2.8\times 10^{-42}$ & $4.8\times 10^{-42}$  \\
$f = 1/2$ & $2.5\times 10^{-45}$ & $5.5\times 10^{-45}$ & $1.5\times 10^{-44}$ & $3.0\times 10^{-44}$ \\
\bottomrule
\end{tabular}
\caption{Best-fit $\sigma_N$ in cm$^2$ at $\delta = 300$~keV. The intervals are given in the Supplemental Material.}
\label{tab:masses}
\end{table}

Figure~\ref{fig:fraction} shows the dependence on $f$. For small $f$ the endothermic case dominates and the interval does not depend on $f$. Above a critical fraction the exothermic case takes over, and the required cross section scales as $1/f$. The critical fraction is $6\times10^{-3}$ at $\delta = 250$~keV, $1.4\times10^{-3}$ at $300$~keV and $3\times10^{-5}$ at $350$~keV. This clarifies the argument of Ref.~\cite{Fan:2026ine}, which treats the two cases separately because comparable contributions would require abundance ratios of $10^{-3}$--$10^{-7}$. Comparable contributions indeed require such ratios. However, for any larger $f$ the combined signal is simply the exothermic one rescaled by $f$. The relevant question is not whether the two cases are comparable, but whether $\chi_2$ survives in more than a tiny fraction.

A benchmark derived for $\chi_1$ alone does not carry over to a halo that also contains $\chi_2$. Consider the thermal pseudo-Dirac benchmark of Ref.~\cite{DiMauro:2026lz}: $m_\chi \simeq 1$~TeV, $\delta \simeq 297$~keV and $\sigma_N \simeq 6.5\times10^{-43}$~cm$^2$. If the halo contains only $\chi_1$ ($f = 0$), this point lies inside the 90\% C.L. interval and explains the LZ event. If half of the halo is $\chi_2$ ($f = 1/2$), the same point is excluded: it lies a factor of $30$ above the upper limit. The fraction of $\chi_2$ in the halo must therefore be considered.

\section{Limitations}
\label{sec:limitations}

The main limitation is the likelihood. The two-dimensional densities of LZ are not public, so we project the data onto three slices in $S1c$ and lose the energy information inside each slice. This lowers our local significance by $0.2$--$0.7\sigma$ and delays the onset of the lower limit from $\simeq 117$ to $140$~keV at $1$~TeV. Two further approximations enter the likelihood. The shape $P_{kj}$ is taken from the $\mathcal{L}_{10}^s$ signal, and it is least accurate at $\delta = 0$, where our upper limit lies 40--50\% above the release, against at most $26\%$ for $\delta \geq 50$~keV. The background is fixed at its post-fit value rather than profiled, but the LZ fit moves every component by at most $1\sigma$.

The recast is validated against LZ at $0.4$, $1$ and $4$~TeV for the endothermic signal. At $10$~TeV it extrapolates. No LZ result exists for the exothermic signal, which we therefore cannot validate directly, although it uses the same nuclear response, halo and likelihood.

All assumptions and their estimated effects are listed in the Supplemental Material. Publication of the two-dimensional densities would remove the main limitation.

\section{Summary}

The intervals published by LZ assume that the halo contains only $\chi_1$. They do not apply when $\chi_2$ also survives. Applied to a halo that contains both states, our recast shows that the exothermic case dominates once the fraction of $\chi_2$ exceeds $10^{-5}$--$10^{-2}$, depending on $\delta$. Above this critical fraction the required cross section scales as $1/f$. Thus, the presence of a subdominant excited-state population cannot in general be neglected when interpreting inelastic direct-detection signals. For comparable populations of the two states, the exothermic channel can reduce the cross section required to account for the LZ candidate by several orders of magnitude and remove the endothermic kinematic cutoff. These results demonstrate that the published LZ interpretation assuming a purely $\chi_1$ halo does not directly apply to scenarios in which both states survive to the present day.

\paragraph{Acknowledgments.} The author thanks João Paulo Pinheiro, David Cabo-Almeida, Francesco Costa and Duarte Feiteira for fruitful and enlightening discussions. The author also thanks António P. Morais and Roman Pasechnik for their supervision and support, and acknowledges financial support from FCT through the doctoral grant PRT/BD/154629/2022.

\bibliographystyle{JHEP}
\bibliography{refs}

\clearpage
\onecolumn
\appendix
\section*{Supplemental Material}
\addcontentsline{toc}{section}{Supplemental Material}
\setcounter{figure}{0}\renewcommand{\thefigure}{S\arabic{figure}}
\setcounter{table}{0}\renewcommand{\thetable}{S\arabic{table}}
\setcounter{equation}{0}\renewcommand{\theequation}{S\arabic{equation}}

\subsection*{A. Nuclear form factor and spectral validation}

This section describes how we build the nuclear form factor of Eq.~\eqref{eq:rate} and how we validate the resulting recoil spectra against those of LZ. The construction has two steps: we distribute the shell-model response among the xenon isotopes, and we correct it to the response used by LZ.

The shell-model response $W_M^{\rm AFH}$ of \texttt{dmdd} is given for natural xenon, i.e.\ summed over isotopes. Eq.~\eqref{eq:rate}, however, needs one form factor $F_i^2$ per isotope, since $v_{\min}$ depends on the nuclear mass $m_A$. We therefore take the Helm form factor of each isotope and multiply it by the common ratio
\begin{equation}
R(E_R) = \frac{W_M^{\rm AFH}(E_R)}{\sum_j \tilde\xi_j A_j^2 F^2_{{\rm Helm},j}(E_R)},
\end{equation}
with $\tilde\xi_j$ the number abundances. The Helm form factor sets only the relative weight of the isotopes. The sum over isotopes returns $W_M^{\rm AFH}$ exactly.

LZ uses the density matrices of Ref.~\cite{Anand:2013yka} with the modifications described in Ref.~\cite{LZ:2023eft}, which are not included in \texttt{dmdd}. We remove the residual difference with the ratio
\begin{equation}
C(E_R) = \left.\frac{(dR/dE_R)^{\rm LZ}}{(dR/dE_R)^{\rm ours}}\right|_{\delta=0,\ m_\chi = 1~{\rm TeV}},
\end{equation}
taken from Fig.~1 of \LZ. It equals $1.00$ at $20$~keV, $1.10$ at $150$~keV and $1.36$ at $248$~keV. Since the differential rate factorizes into a nuclear part and a halo part, $C(E_R)$ fixed at $\delta = 0$ applies to any $\delta$ and $m_\chi$.

\begin{figure}[h]
\centering
\includegraphics[width=0.9\textwidth]{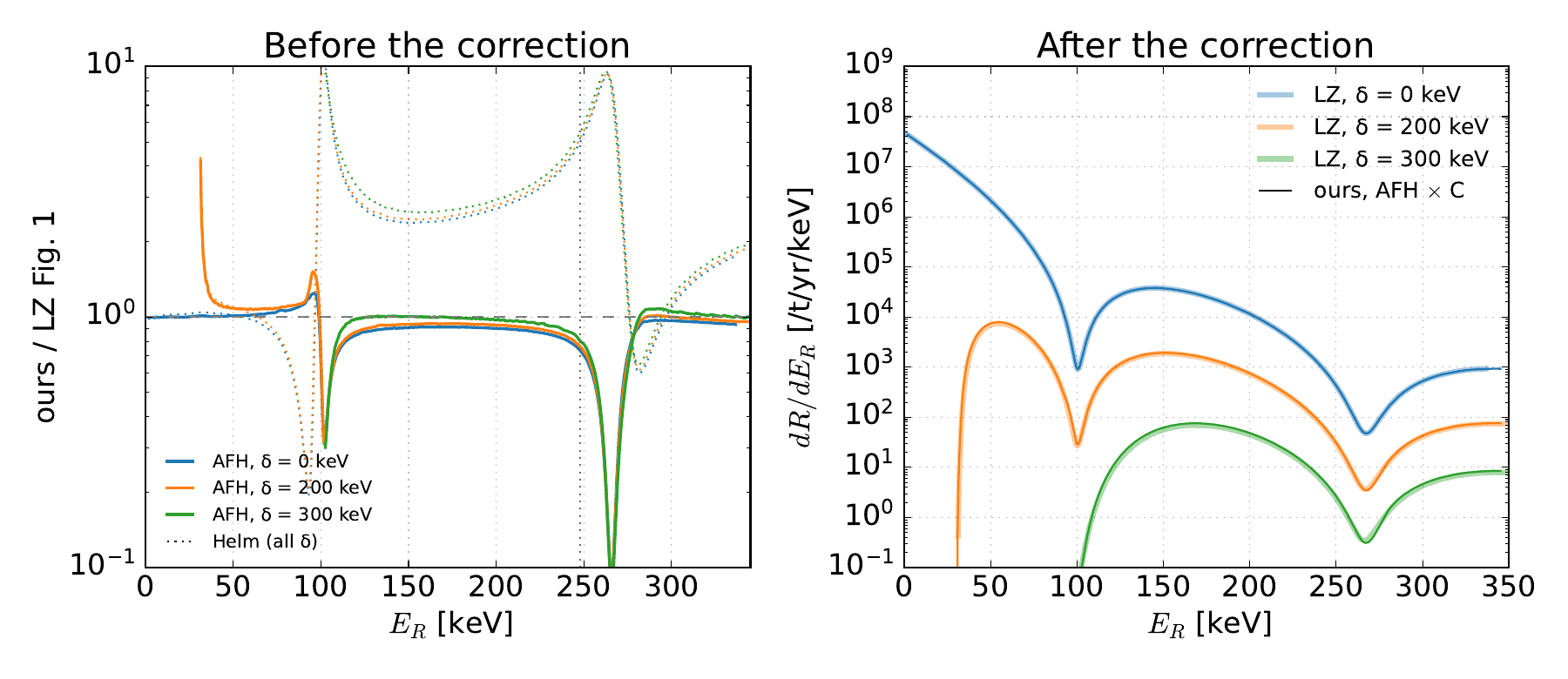}
\caption{Recoil spectra of $\mathcal{O}_1^s$ at $m_\chi = 1$~TeV, with $c_1^s = 1/m_\nu^2$ ($m_\nu = 246.2$~GeV) as in Fig.~1 of \LZ, for $\delta = 0$, $200$ and $300$~keV. Left: ratio of our spectra to those of LZ before the correction $C(E_R)$. Solid lines use the AFH response, dotted lines the Helm form factor. Right: our spectra after the correction (thin) compared with those of LZ (thick).}
\label{fig:spectra}
\end{figure}

The left panel of Fig.~\ref{fig:spectra} compares our spectra with those of LZ before the correction. The normalization agrees at low energy, with ratios $1.00$ at $20$~keV and $1.02$ at $60$~keV. The ratio is nearly the same for $\delta = 0$, $200$ and $300$~keV at a given energy, which tests the halo and the kinematics, since the nuclear response cancels between splittings.  The Helm form factor alone, without the $C(E_R)$, overestimates the LZ spectra by factors of $2.4$ at $150$~keV, $5.0$ at $248$~keV and $8.3$ at $260$~keV.

\subsection*{B. Inputs of the likelihood}

This section details the inputs of the binned likelihood, Eq.~\eqref{eq:likelihood}. We first derive the energy boundaries of the three $S1c$ slices, which set the slice fractions $F_k$. We then describe the signal shape $P_{kj}$, the background $b_{kj}$ and the consistency checks of the extracted data.

The slices are defined in $S1c$, while our spectra are functions of $E_R$. We connect the two through the constant-energy contours of Fig.~4 of \LZ. Along the NR median, the contours of $50$--$300$~keV, in steps of $50$~keV, cross $S1c = 78.8$, $184.0$, $299.0$, $419.5$, $544.1$ and $673.2$~phd. Interpolating between them, the slice edges $S1c = 250$ and $500$~phd and the upper edge of the region of interest, $600$~phd, correspond to $E_R = 128.7$, $232.3$ and $271.7$~keV. The last value coincides with the 50\% point of the efficiency, $269.9$~keV. We cut the spectrum at these energies to obtain $F_k$. For $\mathcal{L}_{10}^s$ this gives $F_k = (0.200, 0.593, 0.206)$, against $(0.208, 0.586, 0.206)$ in Fig.~5 of \LZ. For $\mathcal{O}_1^s$ at $1$~TeV it gives $(0.18, 0.80, 0.013)$ at $\delta = 250$~keV and $(0.035, 0.94, 0.024)$ at $300$~keV. Only about 2\% of the endothermic signal reaches the slice of the candidate, because of the minimum of the nuclear form factor at $265$~keV.

In the bin of the candidate, $P_{3j}$ holds 7.0\% of the signal of the slice, against 4.4\% for a Gaussian band: the NR band has a heavier lower tail. The background in that bin is $3.7\times10^{-5}$ events: accidentals ($1.8\times10^{-5}$), MSSI ($1.5\times10^{-5}$), and neutrinos and neutrons ($0.4\times10^{-5}$). The extracted $\mathcal{L}_{10}^s$ signal sums to $1.047$ events against the fitted $1.0$, and the background to $1597$, $23.4$ and $0.0098$ events in the three slices, against $1593$, $23$ and $1$ observed. Figure~5 of \LZ\ shows $1617$ of the $1710$ events of the science sample; the background model predicts $1621$ in the same window and $1713$ in total.

These comparisons test the extraction, and none of the differences points to an error in it. The values are read from the vector paths of Fig.~5, not digitized by eye, so the reading error is negligible. The signal sum, $1.047$, agrees with the fitted $1.0$ of Table~I of \LZ, which is quoted to one decimal. The differences between background and data, $1597$ against $1593$ and $23.4$ against $23$, are Poisson fluctuations well below $1\sigma$, of the same kind as $1713$ against $1710$ in Table~I. In the highest slice, our background, $0.0098$ events, agrees with the $0.0106 \pm 0.0008$ quoted in the caption of Fig.~5 of \LZ, and the single observed event is the candidate. The remaining $93$ events of the science sample are not shown in Fig.~5, presumably because they lie outside the plotted range, $|x| < 8$, on the ER side. The only loss specific to the figure is that bins below the lower edge of each panel are not drawn, and they carry a negligible fraction of the events.

\begin{figure}[h]
\centering
\includegraphics[width=0.9\textwidth]{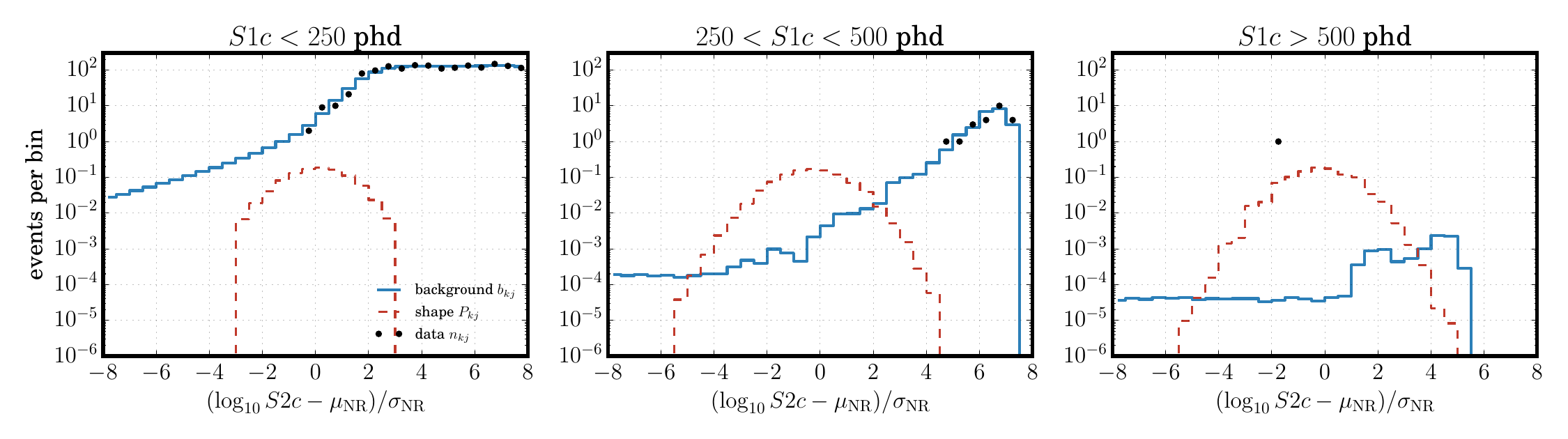}
\caption{Inputs of the binned likelihood, extracted from Fig.~5 of \LZ, in the three $S1c$ slices (left to right) and versus the distance to the NR median, $(\log_{10}S2c - \mu_{\rm NR})/\sigma_{\rm NR}$. Blue: post-fit background $b_{kj}$, in events per bin. Points: observed counts $n_{kj}$. Dashed red: signal shape $P_{kj}$, normalized to unity in each slice (see text). The candidate is the single event at $(\log_{10}S2c - \mu_{\rm NR})/\sigma_{\rm NR} = -1.75$ in the highest slice.}
\label{fig:likinputs}
\end{figure}

Figure~\ref{fig:likinputs} shows the inputs. The dashed curves show the shape $P_{kj}$, normalized to unity within each slice, $\sum_j P_{kj} = 1$. We choose this normalization because $P_{kj}$ is the only signal input taken from LZ. It describes the detector response and is the same for every model, while the slice fractions $F_k$ and the number of events $s$ depend on $m_\chi$, $\delta$ and $f$ and are computed separately. The curves are therefore not the signal curves of Fig.~5 of \LZ, which show the full $\mathcal{L}_{10}^s$ signal, $s\,F_kP_{kj}$ with $s \simeq 1$. The two differ by the slice fraction $F_k$: our curves lie a factor $1/F_k \simeq 5$ above those of LZ in the lowest and highest slices, and $1.7$ above in the middle one. This is a choice of display only. The likelihood, Eq.~\eqref{eq:likelihood}, always uses the full product $sF_kP_{kj}$, and none of our results depends on how $P_{kj}$ is plotted.

\subsection*{C. Statistical details}

This section gives the numerical details of the statistical inference. We describe, in order, the best-fit signal, the $p$-value of each signal hypothesis, the construction of the interval, the comparison with the asymptotic approximation, and the onset of the lower limit.

 The best-fit signal $\hat s$ maximizes $\mathcal{L}(s)$ under $\hat s \geq 0$. It solves $\sum_{kj} n_{kj} a_{kj}/(\hat s a_{kj} + b_{kj}) = \sum_{kj} a_{kj}$, with $a_{kj} = F_k P_{kj}$, when this equation has a positive root, and is zero otherwise. We find the root by bisection, which is fast enough to be applied to every toy.

For each value of $s$, we generate $N_{\rm toys} = 3000$ toys and compute
\begin{equation}
p(s) = \frac{N[\tilde t_s \geq \tilde t_s^{\rm obs}] + 1}{N_{\rm toys} + 1},
\end{equation}
where $N[\tilde t_s \geq \tilde t_s^{\rm obs}]$ is the number of toys with $\tilde t_s$ at least as large as observed. The statistical uncertainty on $p$ near $\alpha$ is $0.005$. We evaluate $p(s)$ on a grid of $70$ values: $0$, $25$ logarithmic values in $[10^{-3}, 1]$ and $44$ linear values in $(1, 12]$. The value $s = 0$ gives $p_0$. The logarithmic part resolves the lower limit, which is a fraction of an event, and the linear part the upper limit, which lies at a few events. The largest upper limit in all our results is $8.6$ events, well below the end of the grid.

We verified the coverage of the construction with a second, independent set of toys, which play the role of the data. For $m_\chi = 1$~TeV, $\delta = 150$ and $300$~keV, and true signals $s = 0$, $0.3$, $1$, $3$ and $6$ events, we generate $5000$ such toys and apply the full procedure to each, with the $p$-value computed from the usual $3000$ toys. The fraction of intervals that contain the true $s$ is $0.90 \pm 0.01$ in all cases, except at $s = 0$ and $\delta = 300$~keV, where it is $0.96$. This overcoverage reflects the discreteness of the Poisson counts in the nearly empty bins of the candidate, and makes the interval conservative.

The 90\% C.L. interval contains all values of $s$ with $p(s) \geq \alpha$. Its limits follow from linear interpolation of $p(s)$ at $\alpha$. The lower limit is zero when $p_0 \geq \alpha$.

\begin{figure}[h]
\centering
\includegraphics[width=0.85\textwidth]{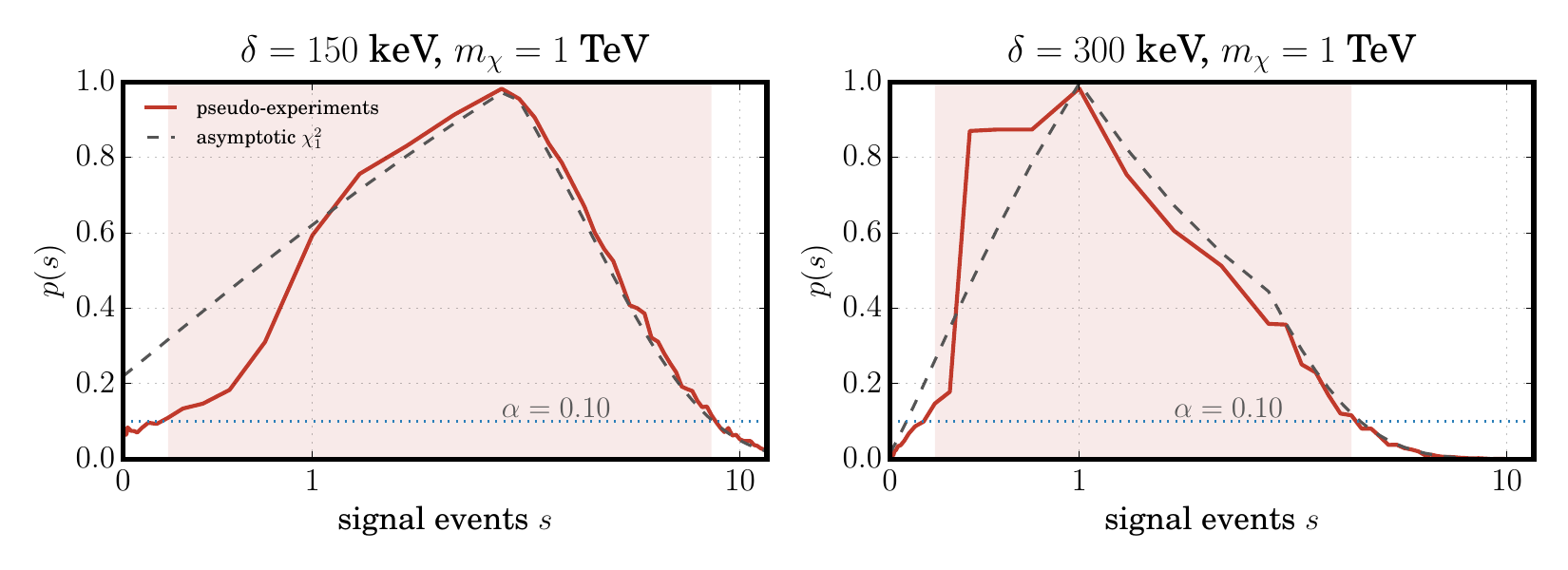}
\caption{$p$-value of the observed $\tilde t_s$ versus the number of signal events $s$, for $m_\chi = 1$~TeV and $\delta = 150$~keV (left) and $300$~keV (right). Solid: toys. Dashed: asymptotic $\chi^2_1$ distribution; at $s = 0$ it is twice the one-sided asymptotic $p_0$ of the main text. Dotted: $\alpha = 0.1$. Shaded: accepted values of $s$, which form the 90\% C.L. interval.}
\label{fig:pvalue}
\end{figure}

Figure~\ref{fig:pvalue} compares $p(s)$ from the toys with the asymptotic $\chi^2_1$ approximation\footnote{Here $\chi^2_1$ denotes the chi-square distribution with one degree of freedom, which $\tilde t_s$ follows asymptotically~\cite{Cowan:2010js}.}. The two agree at large $s$, where many signal events are expected, and differ at small $s$, where the bins of the candidate are nearly empty. The steps in the toy curve at $\delta = 300$~keV reflect the discreteness of the Poisson counts in these bins.

\begin{figure}[h]
\centering
\includegraphics[width=0.6\textwidth]{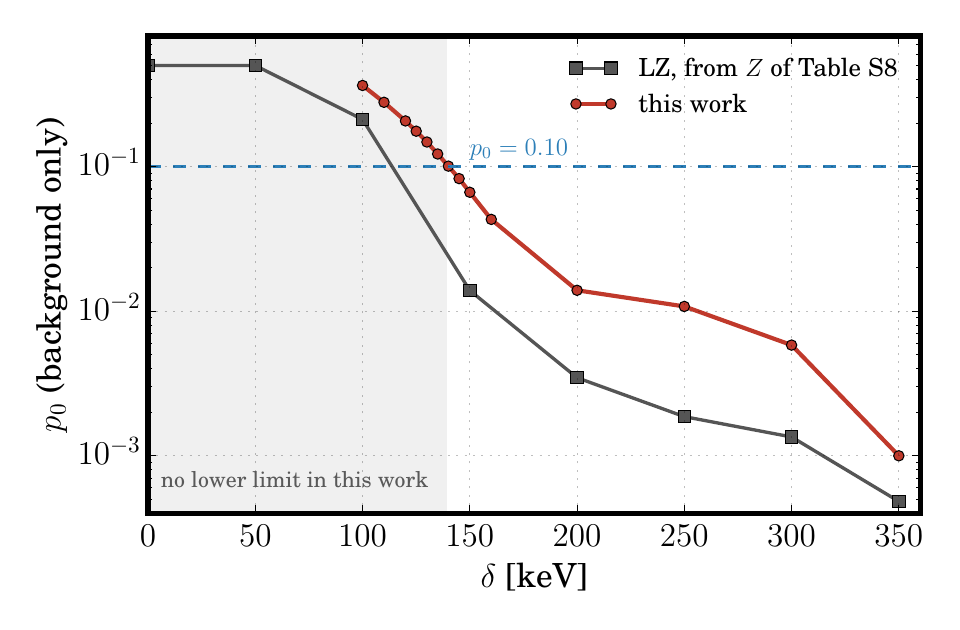}
\caption{Background-only $p$-value $p_0$ versus $\delta$ for endothermic $\mathcal{O}_1^s$ at $m_\chi = 1$~TeV. Red: this work, from $10^5$ toys. Grey: LZ, from the local significances of Table~S8 of \LZ. Dashed: $p_0 = \alpha$; a non-zero lower limit exists below this line. Shaded: region without a lower limit in this work.}
\label{fig:onset}
\end{figure}

 By Eq.~\eqref{eq:onset}, a non-zero lower limit exists where $p_0 < \alpha$. Figure~\ref{fig:onset} shows $p_0$ as a function of $\delta$ at $1$~TeV. Near the onset we scan $\delta$ in steps of $5$--$10$~keV between $100$ and $160$~keV; above $160$~keV we use the grid of the main results. Our lower limit is zero at $\delta = 135$~keV, where $Z = 1.16$, and non-zero at $140$~keV, where $Z = 1.28$. The local significances of Table~S8 of \LZ, converted into $p_0$, cross $\alpha$ near $117$~keV, from linear interpolation of $Z$. Our $p_0$ lies above that of LZ at all $\delta$, which reflects our lower significances, and delays the onset by about $20$~keV.

\clearpage
\subsection*{D. Numerical results}

This section lists the numerical results behind Figs.~\ref{fig:masses} and~\ref{fig:fraction}. Table~\ref{tab:masses_int} gives the 90\% C.L. intervals at $\delta = 300$~keV for the four masses of Table~\ref{tab:masses}, for $\chi_1$ alone ($f = 0$) and for equal abundances ($f = 1/2$). Table~\ref{tab:twocomp} follows the two cases as a function of $\delta$ at $1$~TeV. The last column gives the ratio of the best fits, which measures how much the exothermic case lowers the required cross section. For $f = 0$ there are no entries above $350$~keV, the endothermic cutoff, Eq.~\eqref{eq:deltamax}. At $\delta = 100$~keV the interval for $f = 1/2$ has no lower limit, since $p_0 > \alpha$.

\begin{table}[h]
\centering\small
\begin{tabular}{lccccc}
\toprule
$m_{\chi_1}$ [TeV] & 0.5 & 1 & 2.5 & 5 & 10 \\
\midrule
$f = 0$ & $[0.6,\ 12]\times10^{-42}$ & $[0.4,\ 7.1]\times10^{-42}$ & $[0.5,\ 10]\times10^{-42}$ & $[0.8,\ 17]\times10^{-42}$ & $[1.6,\ 32]\times10^{-42}$ \\
$f = 1/2$ & $[0.4,\ 9.1]\times10^{-45}$ & $[0.8,\ 20]\times10^{-45}$ & $[0.2,\ 5.3]\times10^{-44}$ & $[0.4,\ 11]\times10^{-44}$ & $[0.9,\ 22]\times10^{-44}$ \\
\bottomrule
\end{tabular}
\caption{90\% C.L. intervals on $\sigma_N$ in cm$^2$ at $\delta = 300$~keV.}
\label{tab:masses_int}
\end{table}

\begin{table}[h]
\centering\small
\begin{tabular}{ccccc}
\toprule
$\delta$ [keV] & $f = 0$, best fit & $f = 1/2$, interval & $f = 1/2$, best fit & ratio \\
\midrule
100 & $1.9\times10^{-46}$ & $[0,\ 6.4\times10^{-46}]$ & $5.1\times10^{-47}$ & 0.27 \\
200 & $4.9\times10^{-44}$ & $[3.4\times10^{-46},\ 6.5\times10^{-45}]$ & $1.7\times10^{-45}$ & $3.4\times10^{-2}$ \\
250 & $2.9\times10^{-43}$ & $[5.7\times10^{-46},\ 1.4\times10^{-44}]$ & $3.5\times10^{-45}$ & $1.2\times10^{-2}$ \\
300 & $2.0\times10^{-42}$ & $[8.1\times10^{-46},\ 2.0\times10^{-44}]$ & $5.5\times10^{-45}$ & $2.8\times10^{-3}$ \\
350 & $1.2\times10^{-40}$ & $[1.2\times10^{-45},\ 2.8\times10^{-44}]$ & $7.9\times10^{-45}$ & $6.4\times10^{-5}$ \\
400 & -- & $[2.1\times10^{-45},\ 4.2\times10^{-44}]$ & $1.2\times10^{-44}$ & -- \\
500 & -- & $[8.0\times10^{-45},\ 1.3\times10^{-43}]$ & $3.6\times10^{-44}$ & -- \\
\bottomrule
\end{tabular}
\caption{Cross sections in cm$^2$ at $m_{\chi_1} = 1$~TeV. The ratio is the best fit for $f = 1/2$ over that for $f = 0$.}
\label{tab:twocomp}
\end{table}

\subsection*{E. Assumptions}

The choices made in the recast are listed below, each with its justification and its estimated effect on the results.

\begin{itemize}
\item \textbf{Binned likelihood from Fig.~5 of \LZ:} the two-dimensional densities are not public. It lowers the local significance by $0.2$--$0.7\sigma$ and delays the onset of the lower limit by about $20$~keV.

\item \textbf{Background fixed at its post-fit value:} the LZ fit moves every component by at most $1\sigma$, and the highest slice contains only $0.01$ background events. 

\item \textbf{Science sample only:} the veto samples enter the LZ analysis only through the constraints on the background, which the post-fit values already include. 

\item \textbf{$P_{kj}$ from $\mathcal{L}_{10}^s$:} the detector response depends on $E_R$ only. The upper limit deviates by up to 50\% at $\delta = 0$, and by at most 26\% for $\delta \geq 50$~keV.

\item \textbf{Slice edges from Fig.~4 of \LZ:} the cut reproduces the $\mathcal{L}_{10}^s$ slice fractions of LZ within 1\%. 

\item \textbf{AFH response from \texttt{dmdd}, corrected by $C(E_R)$:} the modified density matrices of LZ are not included in \texttt{dmdd}. The residual difference with the LZ spectra is 9\% at $248$~keV.

\end{itemize}
\end{document}